\documentstyle[aps,amssymb,epsfig,12pt]{revtex}
\begin{document}
\title
{ The Impact of Discoveries of Ferroelectricity and 
\\Charge Disproportionation in Organic Conductors.}
\author{\large{ S. Brazovskii }$^{1,2}$}
\bigskip
\address{
$^1$ Laboratoire de Physique Th\'eorique et de s Mod\`eles Statistiques, CNRS;\\
 B\^at.100, Universit\'e Paris-Sud, 91405 Orsay cedex, France;\\
e-mail:  brazov@ipno.in2p3.fr,  http://ipnweb.in2p3.fr/~lptms/membres/brazov/,}
\address{$^2$ L.D. Landau Institute, Moscow, Russia.}
\date{December 2001
\footnote{Following the talk  at the  Yamada Conference LVI : 
the Fourth International Symposium on 
Crystalline Organic metals, Superconductors and Ferromagnets ISCOM 2001 
\cite{yamada}, published under
another title at the proceedings \cite{yamada-braz}.}
}
\maketitle

\begin{abstract}
Discoveries of the Ferroelectric anomaly (Nad, Monceau, et al) and of the
related charge disproportionation (Brown et al) call for a revaluation of the
phase diagram of the $(TMTTF)_{2}X$ compounds and return the attention
to the interplay of electronic and structural properties. We shall describe a
concept of the Combined Mott-Hubbard state as the source for the
ferroelectricity. We shall demonstrate the existence of two types of spinless
solitons:  $\pi  -$ solitons, the holons, are observed via the activated conductivity; the
noninteger 
$\alpha -$ solitons are responsible for the depolarization of the FE order.
We propose that the (anti) ferroelectricity does exists hiddenly even in
$Se$ subfamily, giving rise to the unexplained yet optical peak.
We remind then the abandoned theory by the author and Yakovenko for the
universal phase diagram which we contrast with the recent one.
\end{abstract}
\newpage

\section{Introduction. Interplay of electronic and structural phases.}

The family $(TMTCF)_{2}X$, with $C=T,S$, demonstrates almost all known
electronic phases, see \cite{jerome}. At higher temperatures $T_{ao}$ there is
also a set of weak structural transitions of "anion orderings" -AOs which are
slight arrangements of chains of counterions X \cite{anions}. At even higher
$T\approx T_{0}$, another \textquotedblleft structureless\textquotedblleft
 transitions where observed sometimes \cite{lawersanne} which origin stayed
mysterious till now. Very recently their nature was elucidated by discoveries
of the huge dielectric anomaly \cite{nad,prl} and of the charge
disproportionation seen by the NMR \cite{brown}. The phase transition has been
interpreted \cite{prl} as the least expected one: to the Ferroelectric (FE)
state. It is triggered by the uniform shift of ions yielding a macroscopic FE
polarization which is gigantically amplified by the charge disproportionation
at the molecular stack. The microscopic theory \cite{prl} for the
\textquotedblright combined Mott-Hubbard state\textquotedblright\ was based on
accounting for two orthogonal contributions, $U_{b}$ and $U_{s}$ to the
Umklapp scattering coming from two symmetry breaking effects: the build in
nonequivalence of bonds and the spontaneous nonequivalence of sites. This
interference resembles the orthogonal mixing in the \textquotedblright
combined Peierls state\textquotedblright\ \cite{matv} in conjugated polymers
of the $(AB)_{x}$ type.

The FE transition in $TMTTF-X$ is a very particular, as well as very bright,
manifestation of the a more general phenomenon of the "charge
disproportionation" or "charge ordering" which becomes recognized as a common
feature of organic conductors (see the review  \cite{fukuyama} and other
talks at this session). Similarly the earliest theoretical prediction
\cite{seo} applies well to a common situation while the pronouncedly $1D$
electronic regime for the observed transition in $(TMTTF)_{2}X$ invokes
particular difficulties as well as allows for a specially efficient treatment
\cite{prl} which is also particularly suited for the FE transition.

The new events call for revaluation of studies in what was thought to be the
abnormal metallic state \cite{jerome}
and return us to the old concept of the universal phase diagram
\cite{physique,jetp} which was based upon effects of interference between
electronic properties and fine symmetry changes due to the AOs or external
fields. (See more in the  Appendix.)

The usual $\mathbf{q}\neq0$ AOs were always observed for \textit{non
centrosymmetric} (NCS) ions, so that the orientational ordering, related to
establishing short contacts of $O$ to $Se$ or $S$ atoms, was supposed to be a
leading mechanism with positional displacements being its consequences only.
But today we should think about a universal mechanism related to the
displacive instability. These two orderings seem to be independent as it is
supported by the latest finding of a sequence of the FE $\mathbf{q}=0$ and
$\mathbf{q}\neq0$ transitions in $(TMTTF)_{2}ReO_{4}$ \cite{nad}. Already
within the nonperturbed crystal structure the anions provoke the
dielectrization. They dimerize intermolecular distances thus doubling the
on-stack unit cell, hence changing the mean electronic occupation from $1/2$
per molecule to $1$ per their dimer. It originates \cite{zagreb} the small
Umklapp scattering $U_{b}$ which opens (according to Dzyaloshinskii and
Larkin, Luther and Emery) the route to the Mott-Hubbard insulator. (See
\cite{braz-00} for a history introduction).

While bonds are dimerized, the molecules stay equivalent above $T_{0}$. The
adventure of the NMR was a clear detection \cite{brown} for appearance at
$T_{0}$ of the site nonequivalence as a sign of the charge disproportionation.
But it brought the same observation both for the structureless transitions and
for the AO already known as a very rare structure of the type 
$\mathbf{q}_{2}=(0,1/2,1/2)$ in $(TMTTF)_{2}SCN$ 
where the site inequivalence was already
known from the structural studies, see \cite{anions}. \emph{{What is the
difference then between the $\mathbf{q}_{2}$ and the \textquotedblright
structureless\textquotedblright\ transitions?}} A clear cut fitting of the
anomaly in $\epsilon(T)$ to the Curie law \cite{nad} suggests that we are
dealing with the least expected case of the Ferroelectric -FE phase. Even more
curiously, it is the FE version of the Mott-Hubbard state which usually is
associated rather with magnetic orderings. The $\mathbf{q}_{2}$ transition
showed already the common polarization along a single stack but alternating in
perpendicular directions which is the anti-FE ordering. For the
\textquotedblright structureless\textquotedblright\ transition all
displacements must be identical $\mathbf{q}=0$, both along and among the
chains/stacks, thus leading fortunately to the FE state. The polar
displacement give rise also to the joint effect of the build-in and the
spontaneous contributions to the dimerization, due to alternations of bonds
and sites. (Both contributions can be of the build in type in the particular
case of the $(TMTSF)_{0.5}(TMTTF)_{0.5}$ mixture \cite{0.5}.)

\section{The combined Mott - Hubbard state and the ferroelectricity.}

\noindent The charge gap $\Delta=\Delta(U)$ appears as interference of two
contributions to the Umklapp scattering: site $U_{s}$\ and bond $U_{b}$, and
it is a function of the total amplitude $U=\sqrt{U_{s}^{2}+U_{b}^{2}}$. The
electronic energy $F_{e}$ depends only on $\Delta$ that is on the total $U$.
But the energy $F_{i}$ of ionic displacements and of the related molecular
distortions depends only on the spontaneous site component $U_{s}$:
$F_{i}=1/2KU_{s}^{2}$. The total energy 
$F_{tot}=F_{e}(U)+1/2KU^{2}-1/2KU_{b}^{2} $ must be minimal over $U$ under the restriction $U>U_{b}$. The
ground state will change if the minimum appears at some $U=U_{0}>U_{b}$ .
Since $U_{0}=U_{0}(T)$ increases with decreasing $T$, there will be a phase
transition at $T=T_{0}$ such that $U_{0}(T_{0})=U_{b}$.

Forms of the Umklapp energy $H$ can be derived from the symmetry alone
\cite{prl}. The non dimerized system allows only for $\sim U_{4}\cos4\varphi$,
which usually is negligibly small as demonstrated by common 4-fold
commensurate CDWs. The site and bond dimerizations originate 
$H_{U}^{s}=-U_{s}\cos2\varphi$ and $H_{U}^{b}=-U_{b}\sin2\varphi$ correspondingly. 
At presence of both types of dimerization the nonlinear Hamiltonian becomes
\[
H_{U}=-U_{s}\cos2\varphi-U_{b}\sin2\varphi=
-U\cos(2\varphi-2\alpha) \, , \ ; \tan2\alpha=U_{b}/U_{s}
\]
For a given $U_{s}$, the ground state is still doubly degenerate between
$\varphi=\alpha$ and $\varphi=\alpha+\pi$ which allows for phase $\pi$
solitons which are the charge $e$ spinless particles, the holons, observed in
conductivity. Also $U_{s}$ itself can change the sign between different
domains of ionic displacements. Then the electronic system must also adjust
its ground state from $\alpha$ to $-\alpha$ or to $\pi-\alpha$, whichever is
closer. Hence the domain boundary $U_{s}\Leftrightarrow-U_{s}$ requires for
the phase soliton of the increment $\delta\varphi=-2\alpha$ or $\pi-2\alpha$
which will concentrate the non integer charge $q=-2\alpha/\pi$ or
$1-2\alpha/\pi$ per chain. While the $3D$ coupled ionic displacements are
described by the MF approach, the electronic degrees of freedom must be
treated exactly at given $U_{s}$ as described in \cite{prl}. The total Umklapp
value is renormalized from $U$ to $U^{\ast}(U)\sim\Delta^{2}=\Delta^{2}(U)$.

There are three contributions to the electronic polarizability:
\newline 1.
\textit{{Ionic displacements}} would already lead to the macroscopic
polarization, like in usual ferroelectrics, but alone they cannot explain the
observed gigantic magnitude of the effect. Their contribution was estimated
\cite{prl} as $\epsilon_{i}\sim10^{1}{T_{0}}/|T-T_{0}| $ which is by $10^{-3}$
below the experimental value $\epsilon\approx2.5\times10^{4}T_{0}/(T-T_{0})$.
\newline 2. 
\textit{{Intergap electronic polarizability}} is regular at
$T_{0}$: $\epsilon_{\Delta}\sim{\omega_{p}^{2}}/{\omega_{t}^{2}}$ where
$\omega_{p}$ is the plasma frequency of the parent metal. This value can be as
large as $\sim10^{4}$ which corresponds indeed to the background upon which
the anomaly at $T_{0}$ is developed.
\newline 3. \textit{{Collective electronic
contribution.}} can be estimated roughly, see more in \cite{prl}, as a product
of above two, $\epsilon_{el}\sim\epsilon_{i}\epsilon_{\Delta}$ which provides
both the correct $T$ dependence and the right order of magnitude of the
effect. We see that the anomalous diverging polarizability is coming from the
electronic system, even if the instability is triggered by the ions which
stabilize the long range $3D$\ FE order.

\section{Discussion, Perspectives and Conclusions.}

By now the revaluation concerns only the $TMTTF$ subfamily which usually, by
the temperature $T_{0}$, is already in the Mott-Hubbard regime. The $TMTSF$
compounds are highly conductive which today does not allow for these difficult
experiments. Nevertheless the transition may be their, just being hidden or
existing in a fluctuational regime like for stripes in High-$T_{c}$ cuprates.
When it is confirmed, then the whole analyses \cite{jerome} of intriguing
abnormal metallic state will be revised as it already should be done for the
$TMTTF$ case. Probably the signature of the FE charge ordered state is already
seen in optical experiments \cite{optics}. Indeed the Drude peak with a
surprisingly low oscillation strength appearing within the pseudogap can be
interpreted now as the optically active mode of the FE polarization; the joint
lattice mass will naturally explain the low weight. Vice versa, the FE mode
must exist in $TMTTF$ compounds where it is probably located below the
optically accessible range of frequencies. Even the optical pseudogap itself
\cite{optics}, being unexpectedly big for $TMTSF$ compounds with their less
pronounced dimerization of bonds, can be enlarged by the spontaneous
dimerization of sites as it was seen explicitly in $TMTTF$ compounds.

\medskip
More generally, there are two conflicting philosophies for these magnificent
materials:
\newline A. The picture of \cite{jerome} implies that the sequence
of electronic phases follows a smooth variation of basic parameters reducible
to the effective pressure, see Fig. 5 in \cite{jerome}. The majority of
compounds with NCS anions were abandoned, presumably their AOs were thought to
exhort ill defined or undesirable complications. The advantages are
appealing:
\newline a) Concentration on simplest examples avoiding structural
effects;
\newline b) Generality in a common frame of strongly correlated
systems driven mostly by basic parameters;
\newline c) Extensive use of
experiments under pressure and by the NMR.
\newline But there is also another
side of the medal:
\newline 1) Concentration on only simplest examples avoiding
the rich information \cite{anions} on correlation of electronic and structural
properties;
\newline 2) Necessity to introduce the case of the NCS anion
$ClO_{4}$ to demonstrate the appearance of the superconductivity under
pressure (The logic of the \textquotedblright effective
pressure\textquotedblright\ demands to show for this compound only the
nonrelaxed phase with the SDW state rather than the relaxed phase where the
superconductivity appears only after the particular structural transition of
the AO.);
\newline 3) Accent upon pressure as a universal parameter and missing
the NMR splitting.
\newline 4) Ignoring the \textquotedblright
structureless\textquotedblright\ transitions which are typical just for these
selected compounds with centrosymmetric anions. (The wide temperature range of
the Curie anomaly in $\epsilon(T)$ tells that the developing of the charge
disproportionation seen as the FE affects a broad temperature region.)

B. The specific picture developed in \cite{physique,jetp,zagreb,osc} suggested
the synthesis of structural and electronic phase transitions with the accent
upon compounds with AOs. It extends naturally to new observations on charge
disproportionation and ferroelectricity. Its main statements are the following
(see \cite{physique} and Ch.6 of \cite{jetp} for applications):
\newline a)
Displacive, rather than orientational, mechanisms are driving the AOs (the
Ernshaw instability of separated charges);
\newline b) Each fine structural
change exerts a symmetrically defined effect which triggers a particular
electronic state;
\newline c) $1D$ \textquotedblright
g-ological\textquotedblright\ phase diagram of the LL results in $2D$, $3D$
phase transitions only with the backup of special symmetry lowering
effects;
\newline d) Main proof for the 1D physics of the Mott transitions is
given by the $\mathbf{q}_{2}$ structure of the $(TMTTF)_{2}SCN$. (Today it is
seen as the charge disproportionation with the anti FE arrangement.)
\newline
e) Superconductivity appears only if the system is drawn away from the half
filling thus avoiding the Mott insulator state. It happens in the relaxed
phase of the $(TMTSF)_{2}ClO_{4}$ thanks to the unique $\mathbf{q}_{3}$ type
AO leading to inequivalence of chains (This is a purely defined case of what
today is called the \textquotedblright internal doping\textquotedblright. Its
magnitude, i.e. the \textquotedblright interchain charge
disproportionation\textquotedblright, was exactly determined from data on fast
magnetic oscillations \cite{osc}.)
\newline But there are major difficulties of
this picture as well.
\newline 1. In applications to $Se$ compounds there are
common problems of any quasi 1D approach confronting the success of the band
picture for the FISDWs (and vice versa!). 
\newline 2. There are cases of the
superconducting state without observation of the particular $\mathbf{q}_{3}$
type of the AO. (Nevertheless, the recent views on independent AOs allows to
suggest that the $\mathbf{q}_{3}$ structure is still their, at least in local
or dynamic form without the long range order. Low $T$ structural studies of,
particularly, $(TMTSF)_{2}PF_{6}$ are required.)

\medskip We finish to say that new events call for a substantial revision of
the contemporary picture of the most intriguing family of organic metals and
for further efforts to integrate various approaches to their studies.

\section{APPENDIX: History Excursions. }

The recent events call again for a unifying  picture of electronic and structural effects which 
returns us to suggestions already made about two decades ago. Below we quote
from publications written in early-mid 80' which views usually have been
ignored by now. 

\begin{center}
\textbf{Extracts from \cite{physique}.} For more details see Ch.6 in
\cite{jetp}.
\end{center}

... We suggest a general model for the phase diagram of the Bechgaard salts in a
way that the variation of electronic states is mainly determined by the
crystal symmetry changes.

\noindent... a complicated phase diagram PD includes  the states:
Metal M, insulator I, Peierls insulator CDW, magnetic (paramagnetic) insulator
MI (PI), antiferromagnetic AF insulator SDW, field induced SDW,
superconductivity SC of singlet or, not excluded yet, triplet  types.

\noindent... 1D divergent susceptibilities give rise to observable
phenomena only if the pair coherence is preserved in the course of the
interchain tunneling. In the gapless regime it can be maintained by proper
interchain electronic phase shifts which can appear due to some symmetry
changes. There are alternating or random potentials for superconductivity and
high magnetic fields for SDW. 

\noindent... This theory permits us to suggest a general
model for the phase diagram of the Bechgaard salts in a way that the variation
of electronic states is mainly determined by the crystal symmetry changes.
The variation of phases follows the change of an anion type $X$, anion
structure, pressure, temperature, magnetic field. 

\noindent... We suggest a simple general model where details of the PD are uniquely 
determined by the anion structure changes. 
Experimental data show us the following correlation between the
anionic structure (characterized by the wave vector $\vec{q}$) and the state of 
the electronic system.

\begin{enumerate}
\item Unperturbed structure. Bonds are dimerized. PD:  
M$\rightarrow$MI$\rightarrow$ SDW. 
\\The last two phases are clearly separated only in $TMTTF$ subfamily.

\item $\vec{q}_{2}=(0,1/2,1/2)$. The molecules are not equivalent. PD:
M$\rightarrow$MI$\rightarrow$SDW (or CDW  = Spin-Peierls). $T_{MI}$\ and
$T_{SDW}$ are well separated ($X=SCN$: $T_{MI}=160K$\ while $T_{SDW}=7K$)

\item $\vec{q}_{3}=(0,1/2,0)$. The neighboring stacks are not equivalent. PD:
M$\rightarrow$SC$\rightarrow$FISDW.

\item $\vec{q}_{4}=(1/2,1/2,1/2)$. The tetramerization. PD: M$\rightarrow$I
transition being driven externally by the AO.
\end{enumerate}

\noindent The rare case 2. helps us to fix the model for the whole family: a
strongly correlated $1D$\ state with the separation of charge- and spin
degrees of freedom. The typical case 1. qualitatively corresponds to the same
model while the separation is less pronounced and interpretation may be
controversial. 

\noindent  The most important for appearance of the SC is the case 3.: the
alternating potentials lead to some redistribution of the charge between the
two types of stacks, hence their system is driven from the two fold
commensurability which removes the Umklapp scattering, destroys the
Mott-Hubbard effect and stabilizes the conducting state down to lower
temperatures where the SC can appear.

\begin{center}
\textbf{\bigskip Extracts from \cite{zagreb}. }
\end{center}

Here are some extracts from \cite{zagreb} which itself was an extension of
earlier observations on effects of counterions in charge transfer CDWs of the
KCP type \cite{kcp}. 

\noindent ...we propose an alternative explanation of $(TMTSF)_{2}PF_{6}$, based on the
fact that this material possesses a weak dimerization gap $\Delta$. This gap
is due to the environment of the given chain, which, unlike the chain itself,
does not posses a screw symmetry along the chain axis. Without the effect of
the environment the band is quarter-filled. The environment ( $PF_{6}$,
etc.) opens a small gap $\Delta$\ in the middle of this band which therefore
becomes half-filled. Hence also small are the corresponding constant for the
Umklapp scattering: $g_{3}\sim g_{1}\Delta /E_{F}$. The effect of $g_{3}$
appears only below sufficiently low temperature 
$T_{3}\sim E_{F}g^{1/2}(g_{3}/g)^{1/g}$, 
 $g=2g_{2}-g_{1}$. 

\noindent ...Assuming the pressure suppresses
$g_{3}$\ and with it $T_{3}$, the Josephson coupling $J$ of superconducting
fluctuation will finally overcome the Umklapp scattering. This interpretation
explains the observations in $(TMTSF)_{2}PF_{6}$ as a result of competition of
the two small (off-chain) parameters, $g_{3}$\ and $J$, rather than as a
result of the accidental cancellation of the large coupling constants $2g_{2}$
and $g_{1}$.

\noindent... In this way there appears a region in the phase diagram where the
superconductivity exists in absence of $g_{3}$, but where the CDW is
introduced by $g_{3}$. 

\noindent... A closer examination of the model shows
that it is the triplet superconductivity.

\end{document}